\documentclass[twocolumn,aps,showpacs,floatfix,prc]{revtex4}
\usepackage[dvips]{epsfig}
\begin{document}
\title{$J/\psi$ suppression in the threshold model and QGP formation time}
\author{A. K. Chaudhuri}
\email[E-mail:]{akc@veccal.ernet.in}
\affiliation{Variable Energy Cyclotron Centre,\\ 1/AF, Bidhan Nagar,
Kolkata 700~064, India}

\begin{abstract}
In the QGP motivated threshold model, 
in addition to the normal  nuclear absorption, $J/\psi$'s are subjected to an additional "anomalous" suppression. We have analysed the recently published PHENIX data on the participant number  dependence  of the
nuclear modification factor for $J/\psi$'s in Au+Au collisions and extracted the anomalous suppression required to explain the data. At mid rapidity $J/\psi$'s are anomalously suppressed only above a threshold density $n_c$=3.73 fm$^{-2}$. The forward rapidity data on the otherhand require that $J/\psi$'s are continuously "anomalously" suppressed. The analysis strongly indicate that in mid rapidity $J/\psi$'s are suppressed in a deconfined medium.
Using the PHENIX data on the participant number dependence of the Bjorken energy density, we have also estimated the QGP formation time. For critical temperature $T_c$=192  MeV, estimated QGP formation time ranges between 0.06-0.08 fm/c.
\end{abstract}  

\maketitle
 
Recently, PHENIX  collaboration  have published their measurements of the centrality dependence of
$J/\psi$ suppression in  Au+Au collisions  at  RHIC  energy,
$\sqrt{s}$=200 GeV \cite{Adare:2006ns}. Data are taken
at mid rapidity ($|y| < .35$) and at forward rapidity ($1.2<|y|<2.2$).  
In most central Au+Au collisions,  
$J/\psi$'s  are more suppressed at forward rapidity than at mid rapidity. Suppression factor is $\sim$ 3 at mid rapidity and $\sim$ 6 at forward rapidity.

There is growing consensus that in central  Au+Au  collisions  at  RHIC, a deconfined state of quarks and gluons (QGP) is produced. It is expected that a deconfined medium, if produced in Au+Au collisions will leave its imprint in $J/\psi$ production. Long back, Matsui and Satz  \cite{Matsui:1986dk}
predicted that in a deconfined medium,  binding
of a $c\bar{c}$  pair  into  a  $J/\psi$  meson will be hindered,
leading to the  so  called  $J/\psi$  suppression  in  heavy  ion
collisions  \cite{Matsui:1986dk} . However, $J/\psi$'s are also suppressed in a nuclear medium. Inelastic interactions of $J/\psi$'s with the nucleons can dissociate them. Suppressed $J/\psi$ production not necessarily imply a deconfined matter formation.
At RHIC energy, it has been further
argued that rather than suppression, $J/\psi$ production will be enhanced
\cite{ Thews:2000rj,Braun-Munzinger:2000px}. 
Due to large initial energy, large number of $c\bar{c}$ pairs will be
produced in initial hard scatterings. Recombination of $c\bar{c}$
can occur, enhancing the charmonium production. However, as mentioned earlier, PHENIX data do not show any indication of $J/\psi$ enhancement. 

PHENIX data on the centrality dependence of $J/\psi$ suppression has been analysed in several models, e.g.
e.g.  comover  model   
\cite{Capella:2000zp}, statistical  coalescence  model    
\cite{Kostyuk:2003kt},  the kinetic model  \cite{Gorenstein:2000ck,Grandchamp:2003uw} or the QCD based nuclear absorption model \cite{Chaudhuri:2006xm}. None of the models give satisfactory description of the experimental data. Recently, we have analysed \cite{Chaudhuri:2006fe} the PHENIX data in the threshold model and found that it do explain the PHENIX mid rapidity data, but not the forward rapidity data. In the present paper, we refine the analysis and extracted the threshold density required to fit the mid rapidity and the forward rapidity PHENIX data. Extracted threshold density is then used to obtain 
a physical parameter, the QGP formation time. We find that QGP formation time is quite small, $\tau \approx 0.06-0.08 fm$. 
 
Blaizot  et  al  \cite{Blaizot:2000ev,Blaizot:1996nq},  proposed the threshold model  to  explain  the NA50 data on  anomalous $J/\psi$ suppression in  158 AGeV Pb+Pb 
collisions at SPS energy \cite{Abreu:2000ni} .  Threshold model tries to mimic the sudden melting of $J/\psi$ in a deconfined medium.  In the model
  fate of a $J/\psi$ depend on the local energy
density. If the energy density exceeds a critical value, the inter quark potential can not bind a $c\bar{c}$ pair into a $J/\psi$. 
It is also assumed that "local" energy density is proportional to 'local' transverse density. Then if the "local" transverse density exceeds a critical or threshold value, deconfined matter is formed and all the $J/\psi$'s are completely destroyed (anomalous suppression). 
One must remember that  
the anomalous suppression is in addition to the
"conventional nuclear absorption".  
The model  neglects the transverse expansion of the system.
It is implicitly assumed that $J/\psi$'s are absorbed
before the transverse expansion sets in.

In the threshold model, number of $J/\psi$ mesons, produced in a AA collision, at impact parameter ${\bf b}$ can be written as,

\begin{eqnarray} \label{eq1}
N^{J/\psi}_{AA}({\bf b}) = &&N^{J/\psi}_{NN} \int  
 d^2{\bf s}  
 T^{eff}_A({\bf s}) T^{eff}_B({\bf b-s}) \nonumber \\
  &&\times S_{anom}({\bf b,s}),
\end{eqnarray}
  
\noindent  where $T^{eff}(b)$ is the effective nuclear thickness,

\begin{equation} \label{eq2}
T^{eff}({\bf  b})=\int_{-\infty}^{\infty}  dz  \rho({\bf   b},z)
exp(-\sigma_{abs}  \int_z^{\infty} dz\prime \rho({\bf
b},z\prime)),
\end{equation}

\noindent $\sigma_{abs}$ being the $J/\psi$-nucleon absorption cross-section.  For the density we use
the Woods-Saxon form.  

\begin{equation}
\rho(r)=\frac{\rho_0}{1+exp((r-R)/a)}, \hspace{.5cm}\int d^3r \rho(r)=A
\end{equation}
 
\noindent with $R$=6.38 fm and $a$=0.535 fm.

$S_{anom}({\bf   b,s})$  in  Eq.\ref{eq1}  is  the  anomalous
suppression factor introduced by Blaizot {\em et al.}  \cite{Blaizot:2000ev,Blaizot:1996nq}. Assuming that  all  the
$J/\psi$'s  get suppressed above a threshold density ($n_c$), the
anomalous suppression can be written as,

\begin{equation}  \label{eq3}
S_{anom}({\bf b,s}) =\Theta (n_c - n({\bf b,s})) 
\end{equation}

\noindent   where  $n_c$ is the critical or the threshold density.
$n({\bf b,s})$   is  the  local transverse  density. It was observed \cite{Blaizot:2000ev} that
by smearing the threshold density by a small amount, one can obtain better fit to the data, but at the expense of an additional parameter ($\lambda$),

\begin{equation}  \label{eq4}
S_{anom}({\bf b,s}) =\frac{1}{2} \left[ 1-\tanh \lambda (n({\bf b,s})-n_c) \right ] 
\end{equation}

Critical ingredient of the threshold model is the "local" transverse density. At impact parameter ${\bf b}$ and at the transverse position ${\bf s}$, local transverse density it can be obtained as,

\begin{eqnarray} \label{eq5}
n({\bf b,s})=&&T_A({\bf s})[1-exp(-\sigma_{NN} T_B({\bf s}-{\bf b}))] \nonumber \\
&&+T_B({\bf b}-{\bf s})[1-exp(-\sigma_{NN} T_A({\bf s}))]
\end{eqnarray}

%%%%%%%%%%%%%%%%%%%%%%%%%%%%%%%%%%%%%
\begin{figure}[ht]
\centerline{\psfig{figure=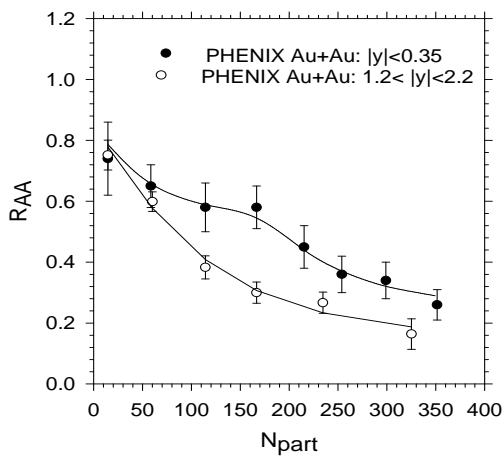,height=10cm,width=8cm}}
\vspace{-3.5cm}   \caption{PHENIX data on the centrality dependence of $J/\psi$ suppression in Au+Au collisions, at mid rapidity and forward rapidity. Best fit to the data in the threshold model is also shown.  }
\label{F1}
\end{figure}
%%%%%%%%%%%%%%%%%%%%%%%%%%%%%%%%%%%%%
With anomalous suppression defined as in Eq.\ref{eq4}, the threshold model have three parameters, $J/\psi$-nucleon absorption cross-section $\sigma_{abs}$, the threshold density $n_c$ and its smearing $\lambda$. 
In the threshold model, $J/\psi$ suppression do not depend explicitly on the rapidity variable. However, experiments do indicate otherwise. In Au+Au collisions, $J/\psi$'s are more suppressed at forward rapidity than at mid rapidity. In the threshold model, such a dependence can only be accommodated if
parameters of the model,  $\sigma_{abs}$,  $n_c$ and  $\lambda$ depend on the rapidity variable. 
We thus 
separately fit the mid rapidity and the forward rapidity PHENIX data to extract those parameters.
Before we proceed further, we would like to note that
it is not unnatural to have rapidity
dependence on the critical parameter $n_c$. For example, it is well known that
the critical temperature of the confinement-deconfinement phase transition depend on the baryon density of the system. Mid-rapidity region is essentially baryon free while at forward rapidity baryon content is non-negiligible. Rapidity dependence of
the critical parameter $n_c$ will then implicitly account for the baryon dependence of the critical parameter.

%%%%%%%%%%%%%%%%%%%%%%%%%%%%%%%%%%%%%
\begin{figure}[ht]
\centerline{\psfig{figure=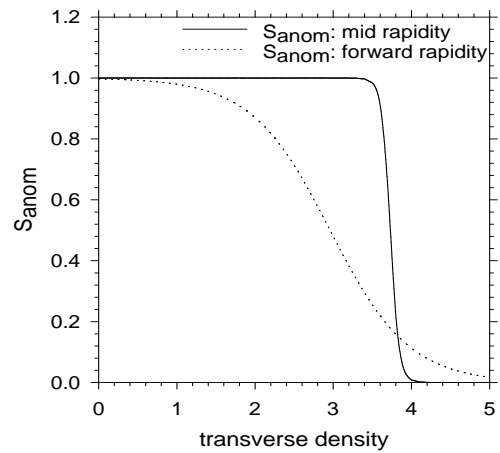,height=10cm,width=8cm}}
\vspace{-3.5cm}   \caption{Anomalous suppression as extracted from the PHENIX data on the centrality dependence of $J/\psi$ suppression. The solid and the dashed lines corresponds to mid rapidity and the forward rapidity data.}
\label{F2}
\end{figure}
%%%%%%%%%%%%%%%%%%%%%%%%%%%%%%%%%%%%%

In Fig.\ref{F1}, PHENIX data are shown. Data points are few and it is not judicious to fit all the three parameters simultaneously.  With the Glauber model of nuclear absorption, we first fit the few peripheral collision (up to $N_{part}$=150) in mid rapidity data and extract the $J/\psi$-nucleon absorption cross-section,   $\sigma_{abs} = 4.39 \pm 0.74 mb$. 
The value is larger than the estimated  
 $\sigma_{abs}$=1-3 mb \cite{Vogt:2005ia} in d+Au collisions.
In peripheral Au+Au collisions, $J/\psi$'s are more
suppressed than in d+Au collisions. With $\sigma_{abs}$ fixed from peripheral collisions, we fit the full data set to find the threshold density $n_c$ and its smearing $\lambda$. Best fit to the mid rapidity data is obtained with
$n_c=3.73 \pm 0.29 fm^{-2}$ and $\lambda=8.96\pm 9.72 fm^2$ .
The solid line in Fig.\ref{F1} shows the fit. The quality of fit is very good.
For the forward rapidity data sets also we use the $\sigma_{abs}$=4.39 mb. As seen in Fig.\ref{F1}, at extreme peripheral collisions, $J/\psi$ suppression in mid and forward rapidity is similar. Best fit to the forward rapidity data set is obtained with   
$n_c=2.96 \pm 0.42 fm^{-2}$ and $\lambda=.99 \pm .94 fm^2$. Here again, as shown in Fig.\ref{F1} the quality of fit is very good. While the threshold model do explain the centrality dependence of $J/\psi$ suppression at mid rapidity as well as at forward rapidity, the anomalous suppression ($S_{anom}$) required for the two data sets are widely different. In Fig.\ref{F2}, we have shown the anomalous suppression $S_{anom}$  as required by the mid and the forward rapidity data.
At mid rapidity, true to the spirit of the threshold model, anomalous suppression shows a step like behavior. At mid rapidity, $J/\psi$ are "anomalously" suppressed only above the threshold transverse density $n_c$=3.73 $fm^{-2}$. But at forward rapidity  $J/\psi$'s are continuously "anomalously" suppressed.
Even though the model fits the data, the spirit  of the model is lost. Step function like anomalous suppression in mid rapidity give strong indication that at mid rapidity, $J/\psi$'s are suppressed in a deconfined medium. Continuous anomalous suppression at forward rapidity on the other hand indicate that $J/\psi$'s are possibly suppressed due a mechanism not related to the confinement-deconfined phase transition.  However at forward rapidity, $J/\psi$ suppression is more complex than envisaged in a simple Glauber like model. Simple Glauber model can not explain either of the two data sets. 

%%%%%%%%%%%%%%%%%%%%%%%%%%%%%%%%%%%%%
\begin{figure}[h]
\centerline{\psfig{figure=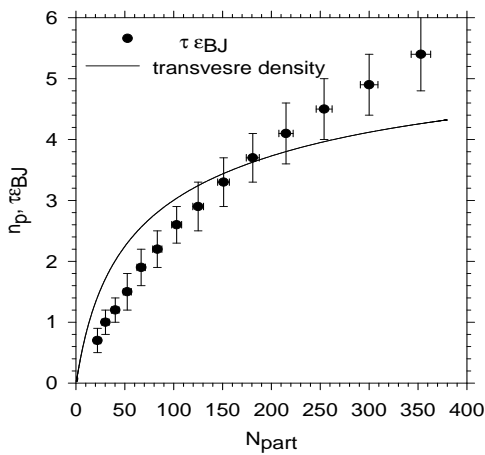,height=10cm,width=8cm}}
\vspace{-3.5cm}   \caption{Solid circles are the PHENIX data on the participant number dependence of Bjorken energy density times QGP formation time ($\tau \varepsilon_{BJ}$) \cite{Adler:2004zn}. The curve is the maximum transverse density that can be achieved in Au+Au collisions.}
\label{F3}
\end{figure}
%%%%%%%%%%%%%%%%%%%%%%%%%%%%%%%%%%%%%

Before we proceed further, we would like to note that  the threshold density as determined here represents the upper limit. Threshold  model neglects some very important effects, e.g. (i) feed back
from $\psi^\prime$ and $\chi$ states and (ii) transverse expansion.
A considerable fraction ($\sim$ 40\%) of $J/\psi$'s are  from decay of $\psi^\prime$ and $\chi$ states  \cite{Satz:2006kb}. That part is completely neglected here. Threshold density for  anomalous suppression of higher states, $\psi^\prime$ and $\chi$ should be less than that for a $J/\psi$. Then presently estimated threshold density $n_c$ represent an upper limit. Additionally,
at RHIC, model studies indicate that in the deconfined phase, the system undergoes significant transverse expansion  \cite{Kolb:2003dz}.
The local transverse density is a key ingredient  to the Threshold model. In an expanding system, local transverse density will be diluted.  
$J/\psi$'s, which   are anomalously suppressed in a static system, may survive in an expanding system due to dilution. Then,  the
presently estimated threshold density will again represent an upper limit.

We now try to connect the estimated threshold density with some physical parameters like threshold energy density or temperature above which $J/\psi$ are anomalously suppressed. As mentioned earlier, threshold density is assumed to be proportional to energy density. If the proportionality factor is known, we can estimate the threshold energy density above which the $J/\psi$'s are anomalously  suppressed.  
As given in Eq.\ref{eq4}, local transverse density is a function of the impact  parameter (${\bf b}$) and the transverse position (${\bf s}$).
For collisions between two identical nucleus at impact parameter ${\bf b}$, maximum transverse density is achieved at the transverse position ${\bf s}={\bf b}/2$. 
In Fig.\ref{F3} we have plotted the transverse density $n_p^{max}({\bf b})=n_p({\bf b},{\bf s}={\bf b}/2)$ as a function of participant number. $n_p^{max}$ increases with the collision centrality. If in a collision with participant number $N_{part}$,
deconfined matter is produced and $J/\psi$'s are anomalously suppressed, at the minimum $n_p^{max}$ 
should exceed the threshold density. As seen from Fig.\ref{F3}, estimated threshold density, $n_c$=3.73 $\pm$ 0.29 $fm^{-2}$ corresponds to  Au+Au collisions with participant number 
$N_{part} = 199.6^{+68.5}_{-56.2}$.

Experimentally one estimate the initially produced energy density by measuring the total transverse energy $E_T$ and using an estimate for the initial reaction volume. In the Bjorken 
model with longitudinal boost-invariance, the energy density is obtained as,  

\begin{equation}\label{eq6}
\varepsilon_{BJ}=\frac{1}{\tau A_T} \frac{dE_T}{dy}
\end{equation}

\noindent where $\tau$ is the formation time, $A_T$ is the overlap area and $dE_T/dy$ transverse energy
per unit rapidity. QGP formation time is an important parameter. Experimental determination of energy density then depends strongly on the estimate of the initial time. 
PHENIX collaboration have measured the transverse energy $E_T$.
Since QGP formation time is not known, they have tabulated the
Bjorken energy density times the formation time as a function of the participant number. 
In Fig.\ref{F3}, we have shown the PHENIX data on the
participant number dependence of the 
$\tau \varepsilon_{BJ}$  \cite{Adler:2004zn}. Like $n_p^{max}$,
$\tau \varepsilon_{BJ}$ increases as the collision centrality increases. PHENIX 
data indicate that a collision with participant number  $N_{part}=199.6^{+68.5}_{-56.2}$, corresponds to   $\tau\varepsilon_{TH}\approx 3.98^{+1.02}_{-1.48}$$GeV fm^{-2}$. $\varepsilon_{TH}$ is the threshold energy density above which
$J/\psi$'s are anomalously suppressed. Corresponding threshold temperature ($T_{TH}$) can be easily obtained using the relation, 
  $\varepsilon=g_{QGP} \frac{\pi^2} {30} T^4$ with $g_{QGP}$=47.5, for a QGP with three flavors.

$T_{TH}$ is the temperature above which $J/\psi$ get dissociated. 
Lattice based potential models indicate that in a deconfined medium, at the critical temperature $T_c$, interquark potential is not sufficiently screened to dissociate $J/\psi$'s. Model calculations indicate that 
$J/\psi$'s can survive upto a temperature of $2.1 T_c$ \cite{Satz:2006kb}. In table \ref{table1}, for a choice of formation time $\tau$, we have tabulated the threshold temperature ($T_{TH}$) and the critical temperature ($T_c$).  For formation time varying between 0.02 fm to 0.2 fm, the critical temperature varies from 150 MeV to 265 MeV. 
Critical temperature for the confinement-deconfinement transition has been accurately estimated in recent lattice calculations, $T_c \sim 192\pm 7\pm4$ MeV \cite{Cheng:2006qk}. As seen from table.\ref{table1}, it corresponds to formation time $\tau$ ranging between 0.06-0.08 fm.  
The time is considerably smaller than the thermal equilibration time $\tau_{eq}\approx 0.6 fm$ \cite{Kolb:2003dz}. QGP is produced early in the collisions.

%%%%%%%%%%%%%%%%%%%%%%%%%%%%%%%%%%%%%
\begin{table}
\caption{\label{table1}Threshold  temperature ($T_{TH}$) 
above which $J/\psi$'s get dissociated and critical temperature ($T_c$) for the confinement-deconfinement transition for  various choices of QGP formation time ($\tau$). }
\begin{ruledtabular}
\begin{tabular}{ccc}
$\tau$ &  $T_{th}$ & $T_c$\\
  (fm)  &   (MeV) & (MeV)\\
\hline
      0.02&     $558.6^{+32.8}_{-61.3}$&     $266.0^{+15.6}_{-29.2}$\\
      0.04&     $469.7^{+27.6}_{-51.5}$&     $223.7^{+13.1}_{-24.5}$\\
      0.06&     $424.4^{+24.9}_{-46.6}$&     $202.1^{+11.9}_{-22.2}$\\
      0.08&     $395.0^{+23.2}_{-43.3}$&     $188.1^{+11.0}_{-20.6}$\\
      0.10&     $373.6^{+21.9}_{-41.0}$&     $177.9^{+10.4}_{-19.5}$\\
      0.12&     $356.9^{+20.9}_{-39.2}$&     $170.0^{+10.0}_{-18.7}$\\
      0.14&     $343.4^{+20.2}_{-37.7}$&     $163.5^{+9.6}_{-17.9}$\\
      0.16&     $332.1^{+19.5}_{-36.5}$&     $158.2^{+9.3}_{-17.4}$\\
      0.18&     $322.5^{+18.9}_{-35.4}$&     $153.6^{+9.0}_{-16.9}$\\
      0.20&     $314.1^{+18.4}_{-34.5}$&     $149.6^{+8.8}_{-16.4}$
\end{tabular}
\end{ruledtabular}
\end{table}
%%%%%%%%%%%%%%%%%%%%%%%%%%%%%%%%%%%

To summarise, in the QGP motivated threshold model, we have analyzed the  PHENIX data on the centrality dependence of $J/\psi$ suppression  in Au+Au collisions. In the threshold model, in addition to the normal nuclear absorption, $J/\psi$'s are   anomalously suppressed, such that, if the local transverse density exceeds a threshold density $n_c$, all the $J/\psi$'s are absorbed. In a careful analysis, we have extracted the threshold
density required to explain the mid rapidity and the forward rapidity PHENIX data. Mid rapidity data are well explained in the model with threshold density $n_c=3.73 \pm 0.29 fm^{-2}$. The data require very small smearing of the threshold density, $\lambda=8.96\pm 9.72 fm$. The forward rapidity data on the other hand require very large smearing, $n_c=2.963 \pm 0.42 fm^{-2}$ and $\lambda=0.99\pm 0.94 fm$. Very large smearing required for the forward rapidity data defeat the essence of the threshold model which tries to mimic the sudden onset of $J/\psi$ in a deconfined medium. We conclude that $J/\psi$ suppression at forward rapidity, though more complex than envisaged in the Glauber model of nuclear absorption,  do not indicate a deconfinement phase transition. $J/\psi$ suppression at mid rapidity which require sudden on set of anomalous suppression above the threshold value $n_c$=3.73 $\pm$ 0.29 $fm^{-2}$, possibly indicate a deconfined matter formation. Using the PHENIX data on participant number dependence of
Bjorken energy density times the formation time,  we have estimated the QGP formation time as $\tau\approx 0.06-0.08 fm$ for critical temperature $T_c\approx=192$ MeV.
  
%%%%%%%%%%%CCCCCCCCCCCCC%%%%%%%%%%%%% 
 

\begin{thebibliography}{99}
%\cite{Adare:2006ns}
\bibitem{Adare:2006ns}
  A.~Adare  [PHENIX Collaboration],
  %``J/psi production vs centrality, transverse momentum, and rapidity in Au +
  %Au collisions at s(NN)**(1/2) = 200-GeV,''
  arXiv:nucl-ex/0611020.
  %%CITATION = NUCL-EX 0611020;%%
%\cite{Matsui:1986dk}
\bibitem{Matsui:1986dk}
T.~Matsui and H.~Satz,
  %``J / Psi Suppression By Quark - Gluon Plasma Formation,''
  Phys.\ Lett.\ B {\bf 178}, 416 (1986).
  %%CITATION = PHLTA,B178,416;%%
%\cite{Thews:2000rj}
\bibitem{Thews:2000rj}
R.~L.~Thews, M.~Schroedter and J.~Rafelski,
%``Enhanced J/psi production in deconfined quark matter,''
Phys.\ Rev.\ C {\bf 63}, 054905 (2001)
[arXiv:hep-ph/0007323].
%%CITATION = HEP-PH 0007323;%%
%\cite{Braun-Munzinger:2000px}
\bibitem{Braun-Munzinger:2000px}
P.~Braun-Munzinger and J.~Stachel,
%``(Non)thermal aspects of charmonium production and a new look at J/psi
%suppression,''
Phys.\ Lett.\ B {\bf 490}, 196 (2000)
[arXiv:nucl-th/0007059].
%%CITATION = NUCL-TH 0007059;%%
%\cite{Capella:2000zp}
\bibitem{Capella:2000zp}
  A.~Capella, E.~G.~Ferreiro and A.~B.~Kaidalov,
  %``Non-saturation of the J/psi suppression at large transverse energy in the
  %comovers approach,''
  Phys.\ Rev.\ Lett.\  {\bf 85}, 2080 (2000)
  [arXiv:hep-ph/0002300].
  %%CITATION = HEP-PH 0002300;%%
%\cite{Kostyuk:2003kt}
\bibitem{Kostyuk:2003kt}
  A.~P.~Kostyuk, M.~I.~Gorenstein, H.~Stoecker and W.~Greiner,
  %``Charm coalescence at RHIC,''
  Phys.\ Rev.\ C {\bf 68}, 041902 (2003)
  [arXiv:hep-ph/0305277].
  %%CITATION = HEP-PH 0305277;%%
%\cite{Gorenstein:2000ck}
\bibitem{Gorenstein:2000ck}
  M.~I.~Gorenstein, A.~P.~Kostyuk, H.~Stoecker and W.~Greiner,
  %``Statistical coalescence model with exact charm conservation,''
  Phys.\ Lett.\ B {\bf 509}, 277 (2001)
  [arXiv:hep-ph/0010148].
  %%CITATION = HEP-PH 0010148;%%
%\cite{Grandchamp:2003uw}
\bibitem{Grandchamp:2003uw}
  L.~Grandchamp, R.~Rapp and G.~E.~Brown,
  %``In-medium effects on charmonium production in heavy-ion collisions,''
  Phys.\ Rev.\ Lett.\  {\bf 92}, 212301 (2004)
  [arXiv:hep-ph/0306077].
  %%CITATION = HEP-PH 0306077;%%
%\cite{Chaudhuri:2006xm}
\bibitem{Chaudhuri:2006xm}
  A.~K.~Chaudhuri,
  %``J/psi production in Au+Au collisions at s**(1/2)(NN) = 200-GeV and the
  %nuclear absorption,''
  Phys.\ Rev.\ C {\bf 74}, 044907 (2006).
  %%CITATION = PHRVA,C74,044907;%%
%\cite{Chaudhuri:2006fe}
\bibitem{Chaudhuri:2006fe}
  A.~K.~Chaudhuri,
  %``J/psi production in Au + Au / Cu + Cu collisions at s(NN)**(1/2) = 200-GeV
  %and the threshold model,''
  arXiv:nucl-th/0610031.
  %%CITATION = NUCL-TH/0610031;%%
\bibitem{Blaizot:2000ev}
J.~P.~Blaizot, M.~Dinh and J.~Y.~Ollitrault,
%``Transverse energy fluctuations and the pattern of J/psi suppression in  Pb Pb
%collisions,''
Phys.\ Rev.\ Lett.\  {\bf 85}, 4012 (2000)
[arXiv:nucl-th/0007020].
%%CITATION = NUCL-TH 0007020;%%
%\cite{Blaizot:1996nq}
\bibitem{Blaizot:1996nq}
  J.~P.~Blaizot and J.~Y.~Ollitrault,
   ``J/Psi Suppression In Pb Pb Collisions: A Hint Of Quark-Gluon Plasma
  %Production?,''
  Phys.\ Rev.\ Lett.\  {\bf 77}, 1703 (1996)
  [arXiv:hep-ph/9606289].
  %%CITATION = HEP-PH 9606289;%%
%\cite{Abreu:2000ni}
\bibitem{Abreu:2000ni}
  M.~C.~Abreu {\it et al.}  [NA50 Collaboration],
  %``Evidence for deconfinement of quarks and gluons from the J/psi  %suppression
  %pattern measured in Pb Pb collisions at the CERN-SPS,''
  Phys.\ Lett.\ B {\bf 477}, 28 (2000).
  %%CITATION = PHLTA,B477,28;%% 
%\cite{Vogt:2005ia}
\bibitem{Vogt:2005ia}
R.~Vogt,
%``Baseline cold matter effects on J/psi production in A A collisions,''
arXiv:nucl-th/0507027.
%%CITATION = NUCL-TH 0507027;%%
%\cite{Satz:2006kb}
\bibitem{Satz:2006kb}
  H.~Satz,
  %``Quarkonium binding and dissociation: The spectral analysis of the QGP,''
  Nucl.\ Phys.\  A {\bf 783}, 249 (2007)
  [arXiv:hep-ph/0609197].
  %%CITATION = NUPHA,A783,249;%%
\bibitem{Kolb:2003dz}
  P.~F.~Kolb and U.~W.~Heinz,
  %``Hydrodynamic description of ultrarelativistic heavy-ion collisions,''
  arXiv:nucl-th/0305084.
  %%CITATION = NUCL-TH 0305084;%%
%\cite{Adler:2004zn}
\bibitem{Adler:2004zn}
  S.~S.~Adler {\it et al.}  [PHENIX Collaboration],
  %``Systematic studies of the centrality and s(NN)**(1/2) dependence of
  %dE(T)/d mu and d N(ch)/d mu in heavy ion collisions at mid-rapidity,''
  Phys.\ Rev.\  C {\bf 71}, 034908 (2005)
  [Erratum-ibid.\  C {\bf 71}, 049901 (2005)]
  [arXiv:nucl-ex/0409015].
  %%CITATION = PHRVA,C71,034908;%%
 %\cite{Kolb:2003dz}
%\cite{Cheng:2006qk}
\bibitem{Cheng:2006qk}
  M.~Cheng {\it et al.},
  %``The transition temperature in QCD,''
  Phys.\ Rev.\  D {\bf 74}, 054507 (2006)
  [arXiv:hep-lat/0608013].
  %%CITATION = PHRVA,D74,054507;%%
\end{thebibliography}
\end{document}